\documentclass[sigconf]{acmart}
\usepackage{epsfig,endnotes,algorithm,verbatim,algpseudocode}

\usepackage{color}
\usepackage{wrapfig}
\usepackage{graphicx}
\usepackage{url}
\usepackage{subcaption}
\usepackage{booktabs}
\usepackage{mwe}
\usepackage{hyperref}
\usepackage{pifont}

\usepackage{times}
\usepackage{datetime}
\usepackage{url}
\usepackage{mathpazo}
\usepackage{tgpagella}

\begin{document}

\acmYear{2017}\copyrightyear{2017}
\setcopyright{rightsretained}
\acmConference{SoCC '17}{September 24--27, 2017}{Santa Clara, CA, USA}
\acmPrice{15.00}
\acmDOI{10.1145/3127479.3132572}
\acmISBN{978-1-4503-5028-0/17/09}

\title{\vspace{-.5cm} \huge{ To Edge or Not to Edge?}}

\author{\vspace{-.65cm} \large{Faria Kalim, Shadi A. Noghabi, Shiv Verma}}
\affiliation{
  \institution{\normalsize{University of Illinois at Urbana-Champaign}, [kalim2, abdolla2, sverma11]@illinois.edu} }


 \begin{abstract}
Edge computing caters to a wide range of use cases from latency sensitive to bandwidth constrained applications. However, the exact specifications of the edge that give the most benefit for each type of application  are still unclear. We investigate the concrete conditions when the edge is feasible, i.e., when users observe performance gains from the edge while costs remain low for the providers, for an application that requires both low latency and high bandwidth: video analytics.
\vspace{-0.5cm}
 \end{abstract}


\maketitle

\vspace*{-1ex}

\vspace*{1ex}
\noindent\textbf{\large Introduction:} Edge computing pushes computation from the cloud to the edge of the network. The concept has recently gained tremendous momentum in both academia and industry  as it can enable a wide range of use cases from latency sensitive to bandwidth/energy constrained, and privacy concerned applications \cite{satyanarayanan2015edge, bonomi2012fog}. Examples include video analytics, online gaming, virtual reality, and self-driving cars.  

Despite the clear potential benefits of the edge, it is unclear what exact specifications the edge should have for its benefits to bloom. This is particularly important for capacity planning (determining how much to invest to see performance gains) and feasibility verification (checking whether the current hardware can meet requirements). Here, we investigate the concrete conditions when an edge is \textit{feasible}, i.e., when users are able to observe performance gains from the edge, at low cost for the service provider. Specifically, we try to answer three main questions: 1) Since locality is key for edge computing, how close should the edge be to the client applications? 2) Do we require as much bandwidth to the edge as we need to the cloud? 3) How much increased load can the  edge sustain? 

\noindent\textbf{The Testbed:} We study the utility of the edge in video analytics applications such as those that analyze traffic patterns near an intersection to warn self-driving cars of passing pedestrians.  Video analytics is usually both bandwidth heavy (generating massive data volumes) and latency sensitive (e.g., acceptable latency is below 185 ms for video in humans \cite{acceptable-latencies} and a few tens of ms for self-driving cars), making the edge a great fit for it. 

Our testbed allows us to vary the key factors that are relevant to our application, i.e., latency, bandwidth and the power of the edge. Thus we can mimic various environments, measure throughput and latency in each case, and determine when  using the edge achieves significant performance gains. 

We built an Android application that captures pictures and performs object detection on them.  We offload object detection to a remote server (edge or cloud) that trains a neural network based model using the YOLO library~\cite{darknet13}, employs it to detect objects in received pictures, and returns them to the client.  

\begin{figure}
\vspace{-0.8ex}
\centering
\includegraphics[width=0.9\columnwidth]{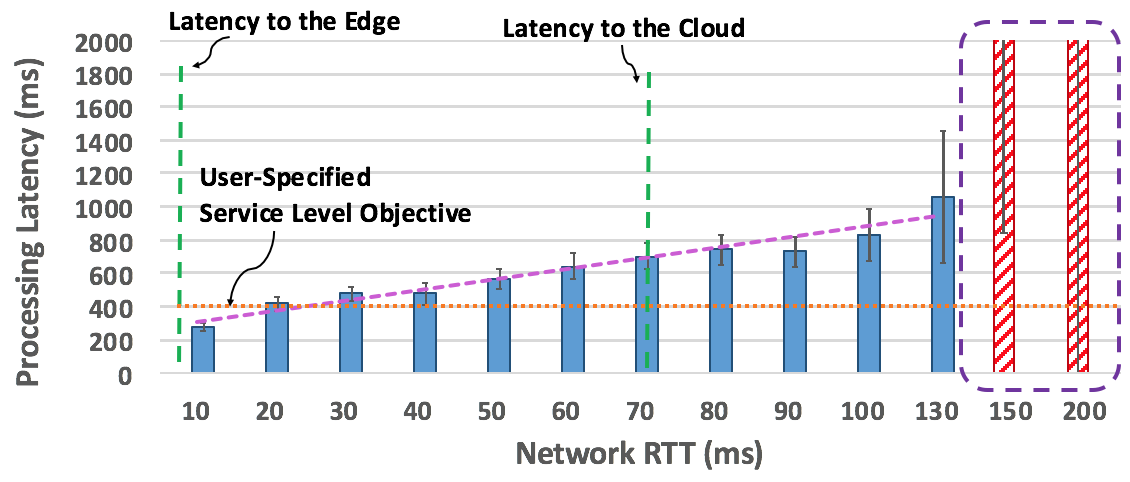}
\vspace{-0.5cm}
\caption{\footnotesize Latency for a processed request increases linearly as network RTT increases. After a threshold, this can cause a build-up of requests at the sender, causing unbounded latencies (shown by red bars).}
\vspace{-0.5cm}
\label{processing-latency}
\end{figure}

\noindent\textbf{Evaluation and Takeaways:} We present one evaluation result in Fig.~\ref{processing-latency}, which shows that even less powerful machines can be used as edge-servers, if the latency to the edge is small. We increase network latency from the client to the edge using Netem. We have a 1 Gbps connection, with an edge provisioned with an NVIDIA Titan X Graphics Card for object detection. Intuitively, the likelihood that latency SLOs will be violated increases in proportion to network RTT. As latencies increase, requests begin to build up at the sender, causing unbounded response latencies. This impacts the cloud more than the edge as it is not only farther away from clients but also has incoming requests from more clients than the edge. 

Our takeaways are: edges can be treated as a resource for a limited number of users who are closest to it. Since this means that there will be a lot less load on the edge, it is possible for resource providers to provision less powerful machines and less bandwidth to the edge. However, they need to ensure that sufficient edges are provisioned such that the network latency to the user is minimized. Additionally, the benefits of the edge truly become obvious when the  computation time (for both the edge and the cloud) is less than the network latency. However, an increase in load affects the edge the same way as the cloud: the network or compute resources can become a bottleneck, causing unbounded latencies. The exact specification of machines and bandwidth vary from application to application. For instance in our setting, an edge placed at 20 ms is ideal. 

In future work, we will consider building on recent work~\cite{alipourfard2017cherrypick} to construct machine learning models that help determine the edge specifications required for a certain number of users and type of workload, while minimizing cost of infrastructure.
\vspace{-2ex}
\bibliographystyle{ACM-Reference-Format}
\bibliography{bib.bib} 
\end{document}